# The Elusive Model of Technology, Media, Social Development, and Financial Sustainability

Aaditeshwar Seth
IIT Delhi and Gram Vaani, India
aseth@cse.iitd.ac.in, aseth@gramvaani.org

*Abstract*—We recount in this essay the decade-long story of Gram Vaani, a social enterprise with a vision to build appropriate ICTs (Information and Communication Technologies) for participatory media in rural and low-income settings, to bring about social development and community empowerment. Other social enterprises will relate to the learning gained and the strategic pivots that Gram Vaani had to undertake to survive and deliver on its mission, while searching for a robust financial sustainability model. While we believe the ideal model still remains elusive, we conclude this essay with an open question about the reason to differentiate between different kinds of enterprises – commercial or social, for-profit or not-for-profit – and argue that all enterprises should have an ethical underpinning to their work.

## I. Background information

Name of Venture: Gram Vaani – http://gramvaani.org

Age of Venture: 10+ years

Size: 70+ full time employees

Management team: Vijay Sai Pratap, Aaditeshwar Seth

Location: India

Form of incorporation: For-profit company

Incubator affiliation: IIT Delhi

Venture fund pipeline: Approximately $600K raised in equity investment from the Media Development Investment Fund and the Indian Angel Network

### A. Problem addressed

Participatory media on the Internet has empowered people by giving them the ability to share knowledge and express themselves openly. Rural and low-income communities in developing regions are however unable to leverage digital participatory media networks for various reasons, including issues of literacy (both educational and technological), affordability of technologies such as smartphones and the Internet, and infrastructure readiness in remote areas. How can these communities be provided with information services despite these challenges?

### B. Role of technology

Gram Vaani operates voice-based participatory media networks through a variety of technology platforms, including community radio [1, 2], IVR (Interactive Voice Response) systems [3, 4, 5], and mobile apps. The use of voice makes the platforms accessible to even poorly literate people, and community radio and IVR systems do not require Internet access either. Gram Vaani's innovation lies in using these simple technologies to build interactive systems that enable people to create and share their own media in voice.

Several impact pathways have evolved through the use of this technology. Information created by the people in a bottom-up manner tends to be more contextual, and improves the knowledge and awareness of the local community [6, 7]. Conversations on the platform about social norms not discussed otherwise leads to an introspection of the issues by the community members, and results in attitude and behavior change [5]. Being able to talk about governance and civic issues on an open platform helps hold powerful stakeholders in check and increases the social accountability of government officials [4]. A specific focus on creating a space for marginalized groups to share their thoughts is extremely empowering for them, especially in a context where such groups have historically been suppressed [8]. Finally, the platforms also provide opportunities for community building through cultural expression, entertainment, and humour [9].

### C. Impact measurement

Over 500,000 users of different Gram Vaani platforms have benefited from socially useful information on health, nutrition, agriculture, livelihood, etc (shown 15-25% improvement in awareness indicators). Over 100,000 users have been directly impacted through improved delivery of government schemes and services in their villages and colonies. Over 2,000,000 users have benefited from access to hyperlocal news and information about their communities.

### D. Learning outcomes

A broad structure is outlined in the following sections, to touch upon our learning our the years at three levels: The strategic challenges of running an enterprise, decisions about alignment with a market based approach or other channels for financial sustainability, and finally some critical reflections for the future.

## II. STRATEGIC CHALLENGES OF RUNNING AN ENTERPRISE

We start with recounting a history of Gram Vaani to set the context. Back in 2007, we were deeply inspired with the success that participatory media had seen in facilitating collective action and governance improvement in developed regions [10, 11], and wanted to find appropriate solutions that could be applied to developing regions where a straightforward transfer of Internet-based participatory media systems would not be suitable [12]. At the same time, we were also cognizant of the poor credibility assurance provided on participatory media platforms [13], and wanted to find a design that would be both participatory as well as able to ensure credible communication. Community radio seemed to be a perfect fit here, and in 2006 India had just announced its new community radio policy permitting non-academic institutions to set up their stations [14], hence it seemed like an ideal medium poised for rapid growth.

Community radio (CR) is a broadcast form of audio-based media, telecast over radiowaves within a radius of 10-15km, with a practice to create local participation in both content development and consumption [15]. These stations are often setup in an economical manner, as a two room setup with one room serving as a content editing and management office space and the other as a recording plus interviewing studio [16]. The studio is sound-proofed in rudimentary ways using egg-trays mounted on walls and thick curtains on the windows. A barebones set of recording equipment consisting of a couple of tripod-mounted microphones in the office, and portable recorders for the field, are used along with a few computers running open-source or even pirated versions of audio editing software. The FM transmitter and antenna tower generally form the largest capital expenditure for the stations. These stations are set up by academic or non-academic non-profit organizations, with a commitment to make the programming inclusive and embedded into the local community. Staff and volunteers from the neighbourhood are trained on audio content production, conducting interviews, coverage of local events, etc [17]. The community is also mobilized to form listener groups to tune-in to the radio broadcasts and discuss the messages. Content needs are determined locally based on what is relevant, with programmes on agricultural advisory, health, education, livelihood, folk songs and entertainment, civic issues, local governance, and other topics [18].

CR stations seemed like an ideal medium to bring participatory media to developing regions, being both participatory in their operations, and with editorial controls to ensure credibility of information. As technologists, after conducting several interactions with stations in India and other countries, we were able to come up with design needs for a technology solution to make the stations more participatory and efficient in their operations [1]. The stations seemed to need an automation system to archive their broadcasts for regulatory compliance, index their content for search and retrieval in the future, stream their content online so that listeners and other stakeholders including donors could listen to the content without physically being in the broadcast station, and most importantly, leverage the increasing penetration of mobile phones to enable listeners to call the station to record their messages, go live on air, talk to an expert over a live conversation, and send and receive SMSes. Such systems existed commercially, but were too expensive to be deployed by small non-profit community groups [19]. Based on a proposal written towards the end of 2007, we won a grant in 2008 to build an open-source radio automation system that could be deployed at CR stations [20].

Further, we realized that CR stations faced a constant challenge of financing their operational expenses. Initial seed grants for capital expenditure and training were able to get the stations started, but operational expenses for staff salaries, equipment upgradation, local travel, etc was an unsolved challenge. To address this challenge, we envisioned also fulfilling the role of an aggregation platform where we could position the potentially nationwide network of stations running our automation software as a syndication platform to run advertisements and sponsored campaigns at a very large scale [21]. Individually it was hard for the stations to market themselves to companies and governments for advertising revenue to reach remote markets, but positioning it as a network we felt could potentially make it more marketable.

Our funding proposal was therefore situated around two objectives [20]. First, to build a radio automation system for CR stations to help them improve their operational efficiency and community engagement. Second, to build a social entrepreneurial business model that would raise advertising revenue from large companies and government departments keen to reach remote regions, and share this revenue with the stations to finance their operational expenses.

However, while our automation product GRINS (Gramin Radio Inter Networking System) released in 2009 was very successful as an appropriate technology solution, our business model did not work out due to external market factors beyond our control. The government which had earlier sounded extremely bullish about having more than 4,000 CR stations around the country, turned out to be slow and demanding in granting licenses for new CR stations. Many non-profit organizations applying for a license had to wait for several years for their application to get vetted. Further, even the small capital requirements for economical CR station setups was not easy to raise by the non-profit organizations, most funding being sectorally aligned rather than towards a broad media platform that could be used by different sectors. Consequently, it was not easy for us to present a large consolidated network to advertisers – a fragmented presence with a CR station partner here and another one there was not convincing enough for advertisers to sign-up.

After two years of having spent growing the GRINS network to more than two dozen stations, we decided to pivot to an alternate medium, although still aligned with our mission of bringing change through participatory media. We built a solution using IVR (Interactive Voice Response) systems, on which people could call through their phones and listen to audio messages or record their own message [22, 23]. Like community radio, this was also geared towards participatory content production, but did not require licenses or any heavy capital expenditure to operate. Being entirely voice-based, it could be accessed even by less literate people. Further, the cost of using the system could be flexibly shifted either to the users to pay or call, or to the organization to run it in a toll-free manner so that users could use it at no cost. A backend web-based system allowed content managers to create and upload their content to be heard on the IVR, or review the content submitted by users and decide whether to publish it on the IVR or not. This system called vAutomate was extended over the years with different applications, such as a discussion forum application as described above where users could listen to or record audio content, a customizable data collection survey application, a case management application to create cases and track the progress made on them, and a helpline application to talk live to a manual operator.

vAutomate was received very well by social development organizations working across different sectors including education, health, agriculture, livelihood, governance, etc. These organizations often had a wide footprint over which they worked and hence the limited geographic reach of CR stations was not adequate for them. They also did not have the time to go through a CR station licensing process. Further, being sectorally aligned themselves, they could choose to use the IVR technology specifically for the use-case in which they were interested. We saw several small projects coming up, ranging from the use of IVR systems to host content for parents about how they could participate in the education of their children, to collection of grievances from urban communities about civic issues, peer-to-peer and expert-based question-answering programmes in agriculture, etc. We envisioned our business model as potentially staying the same, where we would make some revenue through the use of our technology by different partners, and some revenue by running aggregated advertising on the network of IVR instances and CR stations.

As we got deeper into the social sector and began to work with a range of partners, we however realized some new challenges in this space that had become well recognized by now as ICTD (Information and Communication Technologies for Development) [24]. We saw that for the social sector partners to utilize our technology solutions depended a lot on their internal capacity. Some were not able to envision strong use-cases for the technology. Some of them who did, sometimes did not have the capacity to shift the functions that they were currently executing manually to running these functions on the IVR platform. Some of them had not anticipated the effort it would take to create good quality audio content, or to guide and mentor users to record good quality audio content, or to even learn to use IVR systems. We saw many projects ending up as failures in our view, from an impact realization standpoint. Such projects had certainly been impactful to define new use-cases, provide experience to the social sector partners in using ICTs for development, and identify challenges in executing ICTD projects, but due to seemingly avoidable issues that did not require re-learning, we could not help feeling that we could do better to see more impact emerging from our technology innovations.

Towards the end of 2012, one of our self-funded pilots in the state of Jharkhand in using the vAutomate platform for citizen-government engagement, was showing strong promise in emerging as an impactful community empowerment platform [25]. We were working closely with a right-based community organization who would encourage people to record problems that they were facing with government schemes, and the organization would then take up these issues to the concerned officials. These issues included grievances on wage payment delays in government programmes, poor delivery quality of education and health services, non-redressal of civic infrastructural complaints for electricity and water, etc. Upon hearing such problems recorded on a publicly accessible communication platform, many of the problems saw accelerated redressal by government officials. The community based organization also had a deep field presence among highly marginalized groups who did not know how to even use basic technologies like IVR systems, and this taught us the importance of offline training and mentoring of the users for effective use of the technology. Seeing this strong impact, a significant shift upon which we decided this year was to reinvent ourselves from not being just a technology provider in the ICTD space, but to grow this pilot by taking on the end-to-end role of running the participatory media platform ourselves.

We raised an equity investment during 2013 to 2015 around this idea from impact investors, to build a participatory media platform we called Mobile Vaani, in which we would create our own userbase, manage the content ourselves, and build our own institutional partnerships. Like before, we would bring financial sustainability through revenues from advertising and sponsored content hosted on the platform, but unlike the earlier model of being an aggregator of platforms run by different partners using our technology, we would run the platform ourselves. This marked our transformation of moving from a technology company to a media company, and thereby required us to build new functional capability internally. Over the years, we consequently built a creative content creation team, a content moderation team to review user generated content, a community mobilization team to market the platform and train users to use it, a mandatory impact function within the community team, and a business development team to sell advertising and social campaigns on Mobile Vaani. Using the investment, Mobile Vaani soon grew to over 30 districts spread across three states of Jharkhand, Bihar, and Madhya Pradesh, won several awards about its unique technology-based participatory media model, and was mentioned in several lists of innovative startups and social enterprises in India and the world [26]. Organizationally it was a tumultuous phase for the team in formalizing many informal processes, learning to share insights across role-based segregations that began to emerge, meet customer requirements and investor expectations, etc.

We will pick up the thread later for the period beyond 2013, and next reflect on this history to mention some key learning which are likely to be shared with many other enterprises built around technology innovations.

*A. Ecosystem induced pivoting*

We started with working in the community radio space, but the market was not growing fast enough due to slow licensing policies of the government for setting up of new community radio stations. This was in turn constraining our own growth, and two years later we pivoted to an IVR based model for participatory media. Similar experiences have been noted with other ICTs where the policy environment may not have been conducive enough at the time or other barriers may have existed in the ecosystem. The use of drones for drug delivery in remote areas [27], and the use of spectrum whitespaces for remote connectivity [28], are examples of the former, where regulatory caution may have led to slow market growth. It is possible that the challenges we encountered in the slow pace of growth of the CR sector may have also been an effect of regulatory caution on part of the government to not suddenly have lots of autonomous radio stations begin to broadcast content outside of the government control. Another example is that of telecenters [29], where either some parts of the technology stack such as network connectivity were not robust enough for rapid scaling at the time, or a standardized financial sustainability model was non-contextually thrust upon village level entrepreneurs by the government, and resulted in slow or unreliable impact due to ecosystem induced externalities. Enterprises can therefore be expected to pivot due to these external challenges.

### B. Impatience induced pivoting

We found that selling technology to organizations that do not have a culture of using technology, can not only be slow and time-taking, but the ability of the organization to utilize the technology to its potential depends a lot on their own internal capacity and creativity. We saw several of our projects not leading to substantial impact, and ultimately pivoted to start our own participatory media network instead of helping other organizations to set up their networks which we could aggregate. Although we are not able to immediately think of other examples in the social entrepreneurial space that may have gone through a similar phase, some parallels can be drawn with organizational strategies followed by companies like Amazon and Netflix. They started as marketplaces to connect consumers with producers of goods, or content, but have now moved strongly into content production themselves [30]. Probably they saw gaps in the content needs of consumers that were not being met by the traditional content producers, or they wanted to step into the competition themselves, but this is another example of pivoting that enterprises may have to do to remain relevant.

### C. Going beyond technology

We saw that technology alone cannot bring change, it needs supporting pillars around it to nurture its use, guide the users, build services on top of the technology, create institutional linkages to run these services well, etc [3, 31, 32]. Once we started our own participatory media network, we had to quickly expand our team to bring in content specialists, community mobilization experience, and programme design, to support the technology so that it could support the users to use it meaningfully. We realize now that such a diversification beyond technology is not new in the social enterprise space. Social enterprises by definition operate in markets which are not well established, unlike the space occupied by commercial enterprises where substantial functional segregation already exists. For example, if somebody wants to start a new magazine publication for educated urban readers in India, online or offline, then a rich ecosystem already exists of marketing channels to publicize the magazine, networks of columnists who can write for the magazine, advertisers who are looking for new publications to reach out to niche audiences, etc. On the other hand, for a similar initiative in rural markets, Mobile Vaani had to create all these functions internally. Similarly, we have come across many other fascinating social enterprises, such as those creating dairy farmer collectives [33], serving information nuggets to farmers [34], marketing health products for low-income populations [35], etc, all of whom were operating in nascent market ecosystems and had to innovatively diversify beyond their core technology offering.

## III. ORGANIZATIONAL ORIENTATIONS FOR SUSTAINABILITY

We next pick up the thread about how the Mobile Vaani business strategy has evolved since 2013, and continues to evolve, in search of a financial sustainability model for participatory media platforms for rural and low-income users. Although written from the perspective of financing participatory media, the discussion will reveal patterns about the pros and cons of choosing different approaches to create social good: market based, non-profit and charitable, or government driven.

### A. Context of impact investing and a market driven approach

Although the Mobile Vaani model of running the participatory media platform ourselves, gave us a lot of flexibility and focus in bringing substantial impact, the advertising centred business model required equity investment to make the network large enough so that it could begin to attract strong advertising revenues. A small network would not appear attractive enough to large advertisers. Further, brands were still not convinced of taking a digital marketing approach for rural markets. As mentioned earlier, we raised an equity investment during 2013-15 to grow the network, but we have been unable to raise a follow-up impact investment since then. It took us many years but this experience has made us realize a few things about impact investing.

One, we realized that impact investors are exactly what the words mean – they are investors! They invest financial capital, they operate in financial markets, and they expect financial returns. This guides each and every decision that they make. They are patient, no doubt, and we are thankful to them for it. However, they will only invest in an enterprise if there is a clear pathway for financial returns, the scale and the nature of social impact seems to be of secondary importance to them [36]. Two, impact investors are quite risk averse. The volume of capital that they work with, hardly allows for any risk. This, combined with the first point, therefore makes them gravitate towards investing in sectors where successful businesses have already been operating, be it schools or hospitals or even money lending, thus keeping their experimental risk low [37]. They put in money to create similar new businesses, or to bring in existing businesses into the formal economy, so that the enterprises can draw upon new financial capital and expand. This seems to have been the driving force behind MFIs, or social enterprises that operate rural health clinics, or private school chains, in having been able to raise impact investment – all of these businesses were operating in some form or another, and the investment only helped formalize these businesses to draw more external capital.

Our model of creating a participatory media platform with a promise of revenues in the future, seems to have appeared too risky to impact investors [38]. Commercial investors of course invest in these models, which are just like Facebook or Twitter, and possibly have a greater appetite for funding businesses that are creating new markets. Ironically, we have seen commercial investments happening in news and media based enterprises that are not focused on impact at all, but who cater to meet entertainment and self-expression needs of the more up-market segment of people in small towns than of marginalized groups who actually need more support [39].

This brings us to another dilemma that has surfaced for us in this market driven economy. As a for-profit equity investment driven organization, we soon found the winning formula to rapidly expand our userbase at low costs: Hyperlocal news contributed regularly by our team of local volunteers and community reporters, was in great demand and quickly expanded the userbase. However we realized that this formula works reliably only for young men living closer to the urban peripheries, and who already are economically well off. It was not only more costly to reach women users, or poorer families that rely on government schemes, but also their information needs seemed to be more complex than straightforward reporting of local news. We therefore faced a question of how much resources should we spend for this userbase that arguably needs our services more desperately but is more expensive to service [3, 31, 32]. Similarly, being able to support this marginalized userbase to get their rights and entitlements under different government schemes, and to improve the quality of public and private services availed by them, has brought us a lot of social credibility but it required additional effort and hence incurred an additional cost [4]. This raises a question that should we attempt to build social credibility just because we define our vision that way, or is building social credibility also important commercially in a market based economy to gain user trust, and if it is then how much effort should we spend on it?

Note that we have not even touched upon other problems of advertising funded media such as the advertisers being able to influence our agenda. The concerns we have raised are more basic. It remains questionable whether advertising can even become a revenue driver for a participatory media platform that is aimed at rural and low-income communities [40], and if it can then will impact investors be willing to take a bet to try it out? Should a for-profit social enterprise spend its resources to include harder to reach users, or should it constrain itself to the readily accessible users who seem to be better off economically and also of more interest to advertisers? These aspects point towards the need to reflect on what kind of social problems or demographics are likely to gain benefit through market based approaches, and which pockets may get ignored.

The two questions we have raised of the kind of enterprises that impact investors seem to be comfortable with, and the kind of users that for-profit social enterprises may choose to target, are likely to have surfaced for other social enterprises as well. Stories abound such as how farmer collectives working with marginalized tribal groups are unable to raise investment because of their focus on lower income segments and longer time to returns [41], or how innovations in agricultural technology to improve productivity are geared towards large farmers only [42]. This shows that market-based approaches to address social development problems may not be a silver bullet at all, leaving room for other approaches.

We want to add that we have not approached commercial investors as yet, because they do not seem to be interested in the kind of target users that we are reaching out to, and they have also been drawn into the rapidly changing technological landscape of smartphones with digital access leaving the perception that voice-based solutions especially built on IVR systems are outdated. We have recently completed the development of a Mobile Vaani app as well, with evidence of good traction from the users, and we plan to explore this route via commercial investors with a simple user traction based pitch with no mention of impact.

### B. Role of philanthropy

With the Mobile Vaani model stuck in a catch-22 situation of not being large enough to attract substantial advertising revenue, and not able to raise more capital for expansion to become attractive enough for the advertisers, interestingly Gram Vaani has survived through philanthropic grants that use the Mobile Vaani model within different sectors [43, 44, 45]. Programmes funded for social-sector partners to use the Mobile Vaani model, or directly to Gram Vaani to set up new platforms for sector-specific objectives, have brought much needed revenues that have sustained the organization and also the core Mobile Vaani programme although at a slightly diminished scale. The diverse experience gained by the team beyond just the technology development as a result of running Mobile Vaani, has been crucial to execute these programmes with much better success rates than earlier because we are now able to support the partners more extensively on non-technological functions as well, including on content development, community training, and use-case identification.

Due to the sectoral nature of funding in the social development space, although Mobile Vaani has not been able to raise enough grants to do participatory media itself, even this sectorally focused model of funding for health or gender or other programmes has been extremely helpful. It has enabled us to continue making valuable technology enhancements and process improvements, since the underlying technology platform and operational model is the same as that used for Mobile Vaani. The downside however is that we are hardly able to pay attention to Mobile Vaani and the crucial governance-improvement function that it has been serving, because operating as a services organization brings various overheads not necessarily aligned with the original vision of participatory media that we started out with. Most of our funded programmes have been in the space of health, gender rights, financial literacy, etc, with very little on governance, although that was one of the key reasons for us to start working in this space. Similarly, in most programmes we have worked to support partner teams on the ground, and have not been able to replicate the successful models of volunteer collectives that we developed on Mobile Vaani [31]. Although it is revenues from these programmes that ultimately enable us to keep Mobile Vaani running, the programmes now occupy the bulk of our time and mindspace.

It is important to note though that as compared to market-based approaches, donors are indeed keen to reach the most marginalized populations which are not considered attractive by the market as ideal consumers. This is an upside that has helped us retain a focus on rural and low-income communities, and avoid getting sucked into the market driven vortex that gravitates naturally towards seeing eligible users as those who are consumers with disposable income. This also points towards the need to recognize that different user segments may be serviced through different approaches: Market-based approaches for those economically well-off, and grant-funded approaches for poorer people.

There are caveats though in the long term sustainability of philanthropic funding in ICTD. A lot of these ICTD programmes are actually classified as *innovation projects*, that often aim to use new methods (through the use of technology) to tackle known problems. The concept of innovation itself has been argued to having emerged as a means to counter the hegemony of markets, and even of the state in some cases [46]. However the succession to any successful innovation is to transition to a scale-up of the innovation, which can come only from either the markets or the state. This is an inherent contradiction in the approach of many philanthropists, who want to amplify the impact of their investment through the very same means that gave rise to the need for the innovation in the first place. Such a setup then leads to mismatched incentives to displace existing systems, something that is very difficult when dealing with the state. As an example, we successfully piloted two exciting projects related to governance, and in both we engaged closely with the government, but neither was adopted by the state for scale-up. The first was a system that provided voice-based notifications to workers on a national employment rights programme for when their wages were disbursed, and allowed them to dispute it and register a grievance in case of a discrepancy [47, 48]. We took the findings to the central government who went as far as forming a committee but eventually none of the recommendations were taken into consideration, with no transparency about the process. The second was a system that allowed knowledgeable social workers to adopt pending grievances filed by people about welfare entitlements, and assist them to fill any documentation gaps so that the grievances could be processed [4, 49]. Despite strong successes and collaboration from the government in doing the pilot, it was not adopted as a standard process. This makes us believe that innovations which were conceived in the first place to change the status quo, are unlikely to be adopted for the same reason that they would change the status quo which would not be in the interest of several stakeholders.

Markets have been better at embracing disruptive innovations, but as we have discussed earlier, the markets are likely to decide to not service customers who do not have any substantial disposable income to be counted as customers in the first place. All these factors point towards the need to reflect on where grant funds can be deployed in terms of target user segments and sectors of focus, as compared to market-based approaches, but contradictions still remain in this setup and challenge the dominant strategy of philanthropic donors to scale their impact through either the state or the market. We feel this is an unsolved problem so far.

### C. Government support for media platforms

We share an excellent relationship with the government for the Mobile Vaani platform. The platform commands substantial credibility and is regularly used by people to talk about problems they face with government schemes and services, and our volunteers take up these issues to government officials, who often solve them on priority [4]. The officials also actively utilize the platform to send announcements to people, and hence a strong citizen-government engagement relationship is supported through our platforms in a collaborative manner with the government [3]. However, as mentioned earlier, our proposals to the government to fund Mobile Vaani as a participatory media platform aimed at improving governance, have not been successful. On the other hand, some of our sector specific programmes such as to push alerts for better maternal care and child care to families [50, 51], or to track cases of pregnancy, or methods to collect data about activities done by health services staff and cadre [52], have been financed by the government. Many of these programmes also arose via philanthropic grants that catalysed the innovations. This raises a key question about the choice of programmes funded by the government – the evidence so far points towards the state as being willing to fund programmes in low-conflict domains like health, but probably not fund programmes that challenge the status quo of governance, or programmes that help people to get organized to build stronger collectives that might end up challenging the state.

It is possible that our reading of the failure to get government funding for Mobile Vaani may need to be more nuanced though. The government does make extensive payments to media for announcements about schemes and services albeit in a propagandist manner, although they have also used this funding channel as a leverage to thwart media independence much like many commercial advertisers [53]. Our failure to raise funding from the government may therefore be related to our positioning of Mobile Vaani as an empowerment intervention with an apolitical stance, rather than a standard media agency focused on reporting news. While we continue to experiment with different angles, we want to note that even if government funding were to work out, it will not be straightforward to operationalize. Procurement norms, decision making punctuated by the vagaries of electoral priorities, corruption, etc, are significant challenges that make it extremely difficult for small companies to navigate the government system. All this points towards being realistic in assessing what can be expected from the government in supporting media platforms.

### IV. REFLECTIONS FOR THE FUTURE

So far in the discussion above, we took a historical perspective to document some learning about the pros and cons of bringing financial sustainability for Mobile Vaani through different organizational orientations – of being market based, or donor supported, or government supported. Although our discussion has been in the specific context of discovering strategies for participatory media networks, these findings would generalize to other domains as well. We next reflect on the future – the choices we might have going forward, which we debate internally each day.

### A. Unanswered questions – the financing of media

Much of the dilemmas we face about building a financial sustainability model for participatory media, are shared with the broader media industry in general [54, 55]. Although the low level of digital penetration in India has left its media industry largely unscathed so far, but across the developed world digital content distribution platforms such as Facebook have disrupted revenues for content producers. They have drawn critical advertising revenue to themselves, leaving the content producers to resort to methods like paywalls to sustain their operations. This of course constrains the media access to those who have the ability to pay for it, likely making it an unrealistic approach for platforms like Mobile Vaani that are intended to be used by even more marginalized low-income groups. We are toying with the idea of community subscriptions, or financial support provided by local collectives such as women SHG networks or trade unions, where it can be justified that Mobile Vaani is not just providing a participatory media service but also tangible outcomes such as through volunteer driven grievance redressal mechanisms or establishing collective bargaining arrangements [56, 57, 58]. Additionally, we are contemplating crowd-funded programmes where wealthy individual donors can fund participatory media networks for the underprivileged [59]. These mechanisms are currently untested and may provide a viable means for financial sustainability. A hybrid model may be the inevitable eventuality though, to build a combination of different revenue streams: Commercial advertising primarily targeted at middle income consumers and also able to generate a surplus for cross-subsidy of other user segments, donor grants targeted at low-income communities but with a focus on specific sectors only, crowd-funding for the empowerment of marginalized communities, and community subscriptions to support local action aided by the participatory media networks.

### B. To choose one path or not

Our approach so far has been that of pragmatism to sustain our operations by aligning ourselves with different sustainability orientations, but we have also seen that it has constrained us in many ways, in fact drawing our focus away from the governance and empowerment objectives that we started out with. Should we choose just one of these approaches in a do-or-die manner? Or should we continue in the same pragmatic style but formalize the distinction between different orientations by creating subsidiary organizations? Each organization could be aligned towards a specific approach so that it can follow its respective strategy in a focussed manner. For example, a possible categorical distinction could be to have a non-profit that focuses on bringing change through Mobile Vaani for the underserved and raises grants to do this, while a separate for-profit organization focuses on enriching the technology stack to create new product lines that can be offered in a services model to social sector partners. Or should we look at new financial innovations like creating development impact bonds for media platforms? Although these bonds aim to shift the risk towards impact investors, they have also faced criticisms of measurement overhead, the risk of excessive narrowing down of the objectives to make them measurable, and also as a veiled attempt to financialize social services [60]. Further, applying this approach to media may require reliable measurement of media effects, which is a notoriously difficult problem to solve.

Such questions are an intense and exciting source of discussion at Gram Vaani these days, as would be the case also with many other social enterprises trying to figure out their financial sustainability model.

### C. Why does the world need different kinds of enterprises?

In the contemporary context, it increasingly seems that technology driven enterprises, whether explicitly for social good or not, seem to be evolving in a direction of social responsibility. Facebook started out with a strong profit-making and growth focus without paying adequate attention to undesirable outcomes that emerged from its platform, and in the process it lost considerable social credibility. It is now desperately trying to reorient itself [61]. Google started with a motto of do no evil, and is similarly investing heavily in using its platforms for social good to gain both social as well as institutional credibility [62]. Brazen companies like Uber are facing a strong resistance from regulators, and have catapulted the rise of platform cooperatives in many parts of the world to counter social irresponsibility like wage-based exploitation of drivers, tax evasion, and compromising on the social security of drivers [63].

Is the world really becoming a more responsible place by paying attention to ethical concerns that arise from the purely profit-making and shareholder value maximization perspectives of the firm? Floridi suggests that this may is happening due to the increased observability brought about as a result of the information age [64], although media propaganda and rent seeking corporate-government relationships may succeed in countering this [65, 76]. If this is really true though then distinctions for profit-based enterprises Vs non-profit donor funded enterprises, or whether they are doing social good or not, may not be necessary going forward. We may see existing enterprises morph into a new kind of business entity within which different objectives co-exist and regulatory processes along with checks by the employees themselves, ensure appropriate behavior. Some mechanisms through which this could be done is by having a diverse board representation including employee representatives for better corporate governance [66, 67], social audits to take into account the perspective of customers [68], auditing of user demographics and (algorithmic) outcomes to ensure unbiased participation among the userbase [32], profit sharing with employees to reduce wage inequalities [69], and with the public to reduce rentier effects [70], etc. In this case, there may truly just be the need for one kind of a business entity which promotes social good because there may not be any other justifiable ethical goal to work towards. Social enterprises may end up being the only form of enterprises, and the current set of social enterprises may have the opportunity to show the way into the future. We have presented such a framework of the need to design and manage ICT projects with a common underlying ethical system that is followed by the team-members of the project [71]. This ethical system provides crucial guidance in the choice of objectives to pursue, design elements to build, the deployment management practices, and incorporation of feedback from the deployment to shape the design. Ultimately it is the teams working on the project who need to adhere to the ethical system though, and we have discussed why both internal checks imposed by the team-members themselves on one another, and regulatory compliances for enterprises, can help ensure that undesirable outcomes do not arise.

An alternate bleaker path that might however arise is that market based enterprises or even the state could masquerade as working towards social good, but in actuality use that as a veil to mask other agendas. This could result in a shirking space for genuine social enterprises, and this indeed is what seems to be happening in India as more and more large corporate groups are getting into the space of doing social good, with these spaces having been created legitimately by the government [62, 72]. For example, with a stated goal of social good by improving the delivery of welfare services, the Indian state has practically coercively pushed the Aadhaar unique identity system despite documented evidence and heavy resistance of the civil society about the unsuitability of such technology in the Indian context [73, 74, 75]. Other initiatives too like Digital India, or the push towards a cashless economy, are all projected as development initiatives for social good, but adequate attention is not paid to undesirable outcomes that arise from poorly designed policy implementation [65, 76]. Companies like Reliance Jio or PayTM similarly wear the social good badge with the aiming of connecting India or banking the unbanked, and benefit significantly from government policies, but their priorities and inclusion/exclusion criteria are no different from that of any market based enterprise [77]. Another example is Whatsapp which has evaded any regulatory action on the problem of fake news, and has not undertaken any serious efforts itself to address the issues [78]. It seems that politicians, media, and companies alike, not only work together for collaborative gains but also paint an emancipatory and aspirational picture of technology, which drives rapid uptake but often leads to undesirable outcomes [32, 76]. The question this raises then is whether democracies are strong enough for people to be able to shape the political economy to create institutions that can place checks and balances on private and public enterprises alike [79], to ensure that responsible outcomes arise from technology and social good is pursued as their only ethical goal [32, 71]? The answer may be negative at the moment, underscoring the need for both consumers of technologies and for employees who build and manage these technologies, to demand greater social responsibility from enterprises and to eventually convert them into social enterprises [80].

This change we believe can only begin to happen through education and awareness, to understand the ethical guidelines within which technology based enterprises should operate and the goals that they should pursue [71]. We strongly feel that educational institutions need to cultivate capabilities for critical analysis among the future engineers and managers of enterprises about the interactions between technology and society [80]. Once societal foundations are stronger, we are confident that the world will be able to identify the right form that an enterprise should take so that it can create social good in a sustainable manner.

REFERENCES

[1] Koradia, Z., Premi, A., Chandrasekharan, B. and Seth, A. 2012. *Using ICTs to Meet the Operational Needs of Community Radio Stations*. In *Proc.* ACM DEV 2010.

[2] Koradia, Z., Balachandran, C., Dadheech, K., Shivam, M., & Seth, A. 2012. *Experiences of deploying and commercializing a community radio automation system in India*. In *Proc.* ACM DEV.

[3] Moitra, A., Das, V., Gram Vaani, Kumar, A., and Seth, A. 2016. *Design Lessons from Creating a Mobile-based Community Media Platform in Rural India*. In Proc. ICTD

[4] Dipanjan Chakraborty, Mohd Sultan Ahmad, and Aaditeshwar Seth. 2017. *Findings from a Civil Society Mediated and Technology Assisted Grievance Redressal Model in Rural India*. In Proc. ICTD.

[5] Dipanjan Chakraborty, Akshay Gupta, Gram Vaani Team, Aaditeshwar Seth, 2019. *Experiences from a Mobile-based Behaviour Change Campaign on Maternal and Child Nutrition in Rural India*, Proc. ICTD

[6] Seth, A. and Zhang, J. 2008. *A Social Network Based Approach to Personalized Recommendation of Participatory Media Content*, Proc. ICWSM.

[7] Seth, A., Zhang, J., and Cohen, R. 2010. *Bayesian Credibility Modeling for Personalized Recommendation in Participatory Media*. In Proc. UMAP.

[8] Pavarala, V. & Malik, K. 2007. *Other voices: The struggle for community radio in India*. SAGE Publications India.

[9] Koradia, Z., Aggarwal, P., Seth, A., and Luthra, G. 2013. *Gurgaon idol: A singing competition over community radio and IVRS*. In *Proc.* ACM DEV.

[10] OhMyNews. Accessed 2019. *Role in Influencing the Outcome of the 2002 South Korean Presidential Election*. https://en.wikipedia.org/wiki/OhmyNews

[11] McCullagh, D. 2007. *Exiled Journalists Circumvent Censors by Text Messaging*. CNET. https://www.cnet.com/news/exiled-journalists-circumvent-censors-by-text-messaging/

[12] Guo, S., Derakshani, M., Falaki, M.H., Ismail, U., Luk, R., Oliver, E.A., Ur Rahman, S., Seth, A., Zaharia, M.H., and Keshav, S. 2011. *Design and Implementation of the KioskNet System*. Computer Networks.

[13] Seth, A., Zhang, J., & Cohen, R. 2015. *A personalized credibility model for recommending messages in social participatory media environments*. World Wide Web, 18, 1, 111-137.

[14] Ministry of Information and Broadcasting. Accessed 2019. *Guidelines for Community Radio Support Scheme (CRSS)*. https://mib.gov.in/sites/default/files/CRSS_Guidelines.pdf

[15] The Commonwealth of Learning. 2002. *Community-radio Case Studies*. UNESCO.

[16] Ramakrishnan, N. 2007. *CR: A User's Guide to the Technology*. UNESCO.

[17] Tabing, L. 2002. *How to Do Community Radio: A Primer for Community Radio Operators*. UNESCO.

[18] Pringle, I., Mittal, E. and Valdes, M. 2012. *Learning with Community Media: Stories from the Commonwealth and Latin America*. Commonwealth of Learning.

[19] Google. 2008. Google Radio Automation Product Tour. https://www.youtube.com/watch?v=cPQiHyJj7Wk

[20] Seth, A. 2007. *A Proposal to Provide Media Services Rural Areas of India*, Presented at Editor & Publisher and MediaWeek's Interactive Media Conference, Knight News Challenge.

[21] Seth, A. 2010. Community Media Tools for the Bottom of the Pyramid. IEEE COMSNETS Invited talk.

[22] Patel, N., Chittamuru, D., Jain, A., Dave, P., and Parikh, T.S. 2010. *Avaaj Otalo - A Field Study of an Interactive Voice Forum for Small Farmers in Rural India*. In *Proc.* SIGCHI.

[23] CGNet Swara, http://cgnetswara.org/

[24] Series of ICTD (Information and Communication Technologies, and Development) conferences. 2019. https://www.ictdx.org/

[25] Seth, A., Katyal, A., Bhatia, R., Kapoor, D., Balachandra, C., Venkat, V., Moitra, A., Chatterjee, S., Shivam, M., Koradia, Z., and Naidu, P.

2012. *Application of Mobile Phones and Social Media to Improve Grievance Redressal in Public Services*. M4D Web Foundation Workshop.

[26] Economic Times. 2014. 14 Startups to Look Forward to in 2014. https://economictimes.indiatimes.com/biz-entrepreneurship/14-startups-to-look-forward-to-in-2014/slideshow/27778094.cms

[27] Lin. C.A., Shah, K., Mauntel, C. and Shah, S.A. 2018. *Drone Delivery of Medications: Review of the Landscape and Legal Considerations*. American Journal of Health-System Pharmacy, Vol. 75(3).

[28] Aggarwal, V. 2018. DoT Says No to Releasing TV White Space Spectrum, Clarifies it is for Experiments. The Hindu. https://www.thehindubusinessline.com/info-tech/dot-says-no-to-releasing-tv-white-space-spectrum-clarifies-it-is-for-experiments/article8737575.ece

[29] Best, M. and Kumar, R. 2008. *Sustainability Failures of Rural Telecenters: Challenges from the Sustainable Access in Rural India (SARI) Project*. In ITID Volume 4, Number 4.

[30] Yglesias, M. 2013. *Netflix's Race to Become a Content Producer Before the Producers Swallow Netflix*. Slate.com. https://slate.com/business/2013/06/netflix-original-content.html

[31] Moitra, A., Kumar, A., and Seth, A. 2018. *An Analysis of Community Mobilization Strategies of a Voice-based Community Media Platform in Rural India*. Information Technologies & International Development 14 (2018).

[32] Seth, A. 2019. *Ensuring Responsible Outcomes from Technology*. In Proc. COMSNETS, Invited Paper.

[33] Kuttappan, V. 2019. *Personal communication*. https://www.farmery.in/

[34] GSMA. 2014. Mobiles for Development: Case study – Reuters Market Light. https://www.gsma.com/mobilefordevelopment/resources/reuters-market-light/

[35] Radjou, N., Prabhu, J. and Ahuja, S. 2012. *When Ingenuity Saves Lives: Embrace Infant Warmers*. Harvard Business Review.

[36] Prominent impact investor. 2016. *Personal communication*, paraphrasing response when asked about whether their MFI investees were indeed having impact and providing customer satisfaction: "We leave impact measurement to the entrepreneur, in whatever way they want to define it. We do not impose any impact reporting requirements".

[37] Pandit, V. and Tamhane, T. 2017. *Impact Investing: Purpose-driven Finance Finds Its Place in India*. Mckinsey&Company report.

[38] Prominent impact investor. 2018. *Personal communication*, paraphrasing when asked about whether they could support the app development for Mobile Vaani, given that the IVR version has had a lot of traction and impact: "We cannot assume that the Mobile Vaani app will take off because the IVR did well. You should build the app and validate its traction through standard metrics of user engagement".

[39] Kashyap, S. 2018. If 2018 Showered Love on Bharat, What Will 2019 Bring for Indian Languages. Yourstory. https://yourstory.com/2019/01/2018-bharat-india-startup-languages/

[40] Caribou Digital. 2019. *Paying Attention to the Poor: Digital Advertising in Emerging Markets*.

[41] Bhartia, A. 2019. Where do the Disruptors Go? IDR Online. https://idronline.org/where-do-the-disruptors-go/

[42] Fleming, A., Jakku, E., Lim-Camacho, L., Taylor, B., and Thorburn, P. 2018. *Is Big-Data for Big Farming or for Everyone? Perceptions in the Australian Grains Industry*. Agronomy for Sustainable Development.

[43] Gram Vaani. Accessed 2019. *Building Healthy and Equitable Societies*. https://gramvaani.org/?p=2929

[44] Gram Vaani. Enhancing Access to Livelihood Opportunities. https://gramvaani.org/?p=3022

[45] Gram Vaani. Building Resilience and Ensuring Access to Safety Nets. https://gramvaani.org/?p=3151

[46] Aoyama, Y. and Parthasarathy, B. 2016. *The Rise of the Hybrid Domain: Collaborative Governance for Social Innovation*. Edward Elgar Publishing.

[47] Srinivasan, V., Vardhan, V., Kar, S., Asthana, S., Narayan, R., Singh, P., Chakraborty, D., Singh, A. and Seth, A. 2013. *Airavat: An Automated System to Increase Transparency and Accountability in Social Welfare Schemes in India*. In Proc. ICTD.

[48] Chakraborty, D. and Seth, A. 2015. *Building Citizen Engagement into the Implementation of Welfare Schemes in Rural India*. In Proc. ICTD.

[49] Gram Vaani. 2017. *Integrated Model for Grievance-Redressal and Access to Justice*. UNDP final project report.

[50] Viswanathan, V. 2017. *JEEViKA Mobile Vaani: Encouraging Women to Access Maternal Health Information Through Mobile Phones*. Gram Vaani. https://gramvaani.org/?p=3118

[51] Digital Impact Alliance. 2018. Beyond Scale: How to Make Your Digital Development Program Sustainable. https://www.rethink1000days.org/publications/beyond-scale-how-to-make-your-digital-development-program-sustainable/

[52] BIRAC. 2018. *Immunication Data: Innovating for Action (IDIA) Grand Challenges India – Gram Vaani's proposal: Image Recognition Based Data Entry Processes to Ensure Immunization Completeness and Auditing of Reported Data*.

[53] Scroll staff. 2019. Centre Freezes Advertisements to the Times Group, ABP Group, and the Hindu: Report. https://scroll.in/latest/928790/centre-freezes-advertisements-to-the-times-group-abp-group-and-the-hindu-report

[54] Robinson, J.J., Grennan, K. and Schiffrin, A. 2015. Publishing for Peanuts: Innovation and the Journalism Start-up. Columbia University. https://www.cima.ned.org/wp-content/uploads/2015/11/PublishingforPeanuts.pdf

[55] Schiffrin, A. 2019. *Fighting for Survival: Media Startups in the Global South*. Columbia University. https://www.cima.ned.org/resource/fighting-for-survival/

[56] C&A Foundation Staff. 2019. Connecting for Collective Action. https://www.candafoundation.org/latest/stories/working-conditions/connecting-for-collective-action

[57] Nagaraj, A. 2019. Thousands of Indian Factories Under Scrutiy Over 'Miserable' Conditions. Thomson Reuters Foundation. https://in.reuters.com/article/india-workers-rights/thousands-of-indian-factories-under-scrutiny-over-miserable-conditions-idINKCN1S51DM

[58] Ruthven, O. 2018. Labour Reform is Fine but Who Holds Employers to Account When Government Fails? The Wire. https://thewire.in/labour/rights-at-work-who-holds-employers-to-account-when-the-government-fails

[59] Indie Voices and Gram Vaani. Accessed 2019. *Crowd-fund Audio Journalism*. https://gramvaani.org/?p=2265

[60] McKay, K. 2013. *Debunking the Myths Behind Social Impact Bond Speculation*. Stanford Innovation Review.

[61] Thompson, N. and Vogelstein, F. 2018. *Inside the Two Years that Shook Facebook – And the World*. Wired Magazine https://www.wired.com/story/inside-facebook-mark-zuckerberg-2-years-of-hell/.

[62] Indian Railways and Google. Accessed 2019. *Station WiFi Project: Bringing Internet to Millions of Passengers in Railway Stations*. https://www.railwire.co.in/station-wi-fi-project.php

[63] Wilemme, G. 2017. Regulating Uber. Pasis Innovation Review. http://parisinnovationreview.com/articles-en/regulating-uber

[64] Floridi, L. 2010. *Ethics After the Information Revolution (Chapter 1)*. Cambridge University Press.

[65] Sen, A., Agarwal, A., Guru, A., Choudhuri, A., Singh, G., Mohammed, I., Goyal, J., Mittal, K., Singh, M., Goel, M., Gupta, S., Pathak, S., Madapur, V. and Seth, A. 2018. *Leveraging Web-data to Monitor Changes in Corporate-Government Interlocks in India*. In Proc. ACM COMPASS.

[66] Justin Fox. 2018. *Why German Corporate Boards Include Workers*. http://www.bloomberg.com/view/articles/2018-08-24/why-german-corporate-boards-include-workers-for-co-determination

[67] Mike Cooley and Frances O'Grady. *Architect or Bee?: 2016. The Human Price of Technology*. Spokesman Books, England.

[68] McRitchie, J. *A More Cooperative Based Twitter*. Accessed 2019. https://www.corpgov.net/2017/05/a-more-cooperative-based-twitter/.

[69] Neal Gorelflow. 2018. *What if Uber was owned and governed by its drivers*?, http://evonomics.com/uber-sharing-economys-biggest-threat/

[70] Blasi, J. and conway, M. 2018. *A Better Way to Share the Wealth*. Politico. https://www.politico.com/agenda/story/2018/11/20/wealth-inequality-policy-solutions-000790

[71] Seth, A. 2019. *Ethical Underpinnings in the Design and Deployment of ICT Projects*. Working paper.

[72] Tata Trusts and Google. Accessed 2019. Internet Saathi: Bridging the Online Gender Divide in India. https://internetsaathiindia.org/

[73] Jean Drèze. 2017. *Sense and Solidarity: Jholawala Economics for Everyone*. Permanent Black.

[74] Reetika Khera. 2017. *Impact of Aadhaar in Welfare Programmes*. In Economic and Political Weekly.

[75] Singh, S.S. Accessed 2019. Death by Digital Exclusion? On the Faulty Public Distribution System in Jharkhand. The Hindu. https://www.thehindu.com/news/national/other-states/death-by-digital-exclusion/article28414768.ece

[76] A. Sen, Priya, P. Aggarwal, S. Verma, D. Ghatak, P. Kumari, M. Singh, A. Guru, and A. Seth. 2019. *An Attempt at Using Mass Media Data to Analyze the Political Economy Around Some Key ICTD Policies in India*. In Proc. ICTD.

[77] S. Radhakrishnan. 2015. *"Low Profile" or Entrepreneurial? Gender, Class, and Cultural Adaptation in the Global Microfinance Industry*, World Development Vol. 74.

[78] Ponniah, K. 2019. WhatsApp: The 'Black-hole' of Fake News in India's Elections. BBC News. https://www.bbc.com/news/world-asia-india-47797151

[79] A. Sen, D. Ghatak, K. Kumar, G. Khanuja, D. Bansal, M. Gupta, K. Rekha, S. Bhogale, P. Trivedi, and A. Seth. 2019. *Studying the Discourse on Economic Policies in India Using Mass Media, Social Media, and the Parliamentary Question Hour Data*. In Proc. ACM COMPASS.

[80] A. Seth. 2018. *Course outline: Ethics in Applied CS*, Department of Computer Science and Engineering, and the School of Information Technology, Indian Institute of Technology Delhi, http://act4d.iitd.ernet.in/act4d/index.php?option=com_content&view=article&id=37&Itemid=45